\documentclass{article}
\usepackage[final]{neurips_2022}

\usepackage{stfloats}
\usepackage{graphicx}
\usepackage{amsmath}
\usepackage{amssymb}
\usepackage{booktabs}
\usepackage[pagebackref,breaklinks,colorlinks]{hyperref}
\hypersetup{
    colorlinks=true,
    linkcolor=blue,    
    citecolor=blue,     
    urlcolor=blue      
}
\usepackage{longtable}
\usepackage[capitalize]{cleveref}
\usepackage{makecell}
\usepackage{multirow}
\usepackage{pifont}
\usepackage[numbers]{natbib}
\usepackage{caption}
\usepackage{subcaption}

\setcitestyle{square}
\setcitestyle{citesep={,}}

\crefname{section}{Sec.}{Secs.}
\Crefname{section}{Section}{Sections}
\Crefname{table}{Table}{Tables}
\crefname{table}{Tab.}{Tabs.}

\title{SA-Med2D-20M Dataset: Segment Anything in 2D Medical Imaging with 20 Million masks}

\author{
Jin Ye\textsuperscript{1}\thanks{Equal contribution. $^\dagger$ Corresponding author.}
\qquad Junlong Cheng\textsuperscript{1,2}\footnotemark[1]
\qquad Jianpin Chen\textsuperscript{1}
\qquad Zhongying Deng\textsuperscript{1}
\qquad \textbf{Tianbin Li\textsuperscript{1}}
\\
\textbf{Haoyu Wang\textsuperscript{1}}
\qquad \textbf{Yanzhou Su\textsuperscript{1}}
\qquad \textbf{Ziyan Huang\textsuperscript{1}}
\qquad \textbf{Jilong Chen\textsuperscript{2}}
\qquad \textbf{Lei Jiang\textsuperscript{2}}
\\
\textbf{Hui Sun\textsuperscript{1}}
\qquad \textbf{Min Zhu\textsuperscript{2}}
\qquad \textbf{Shaoting Zhang\textsuperscript{1}}
\qquad \textbf{Junjun He\textsuperscript{1, $\dagger$}}
\qquad \textbf{Yu Qiao\textsuperscript{1, $\dagger$}}\\
\\
\textsuperscript{1}Shanghai AI Laboratory \qquad
\textsuperscript{2}Sichuan University \\ 
{\tt\small \{yejin, chengjunlong, zhangshaoting, hejunjun, qiaoyu\}@pjlab.org.cn}
}

\begin{document}

\maketitle

\begin{abstract}
Segment Anything Model (SAM) has achieved impressive results for natural image segmentation with input prompts such as points and bounding boxes. Its success largely owes to massive labeled training data. However, directly applying SAM to medical image segmentation cannot perform well because SAM lacks medical knowledge --- it does not use medical images for training. To incorporate medical knowledge into SAM, we introduce SA-Med2D-20M, a large-scale segmentation dataset of 2D medical images built upon numerous public and private datasets. It consists of 4.6 million 2D medical images and 19.7 million corresponding masks, covering almost the whole body and showing significant diversity. This paper describes all the datasets collected in SA-Med2D-20M and details how to process these datasets. Furthermore, comprehensive statistics of SA-Med2D-20M are presented to facilitate the better use of our dataset, which can help the researchers build medical vision foundation models or apply their models to downstream medical applications. We hope that the large scale and diversity of SA-Med2D-20M can be leveraged to develop medical artificial intelligence for enhancing diagnosis, medical image analysis, knowledge sharing, and education. The data with the redistribution license is publicly available at \url{https://github.com/OpenGVLab/SAM-Med2D}.
\end{abstract}

\section{Introduction}



\begin{figure}[t]
  \centering
  \begin{subfigure}[b]{1.0\textwidth}
    \centering
    \includegraphics[width=1.0\textwidth]{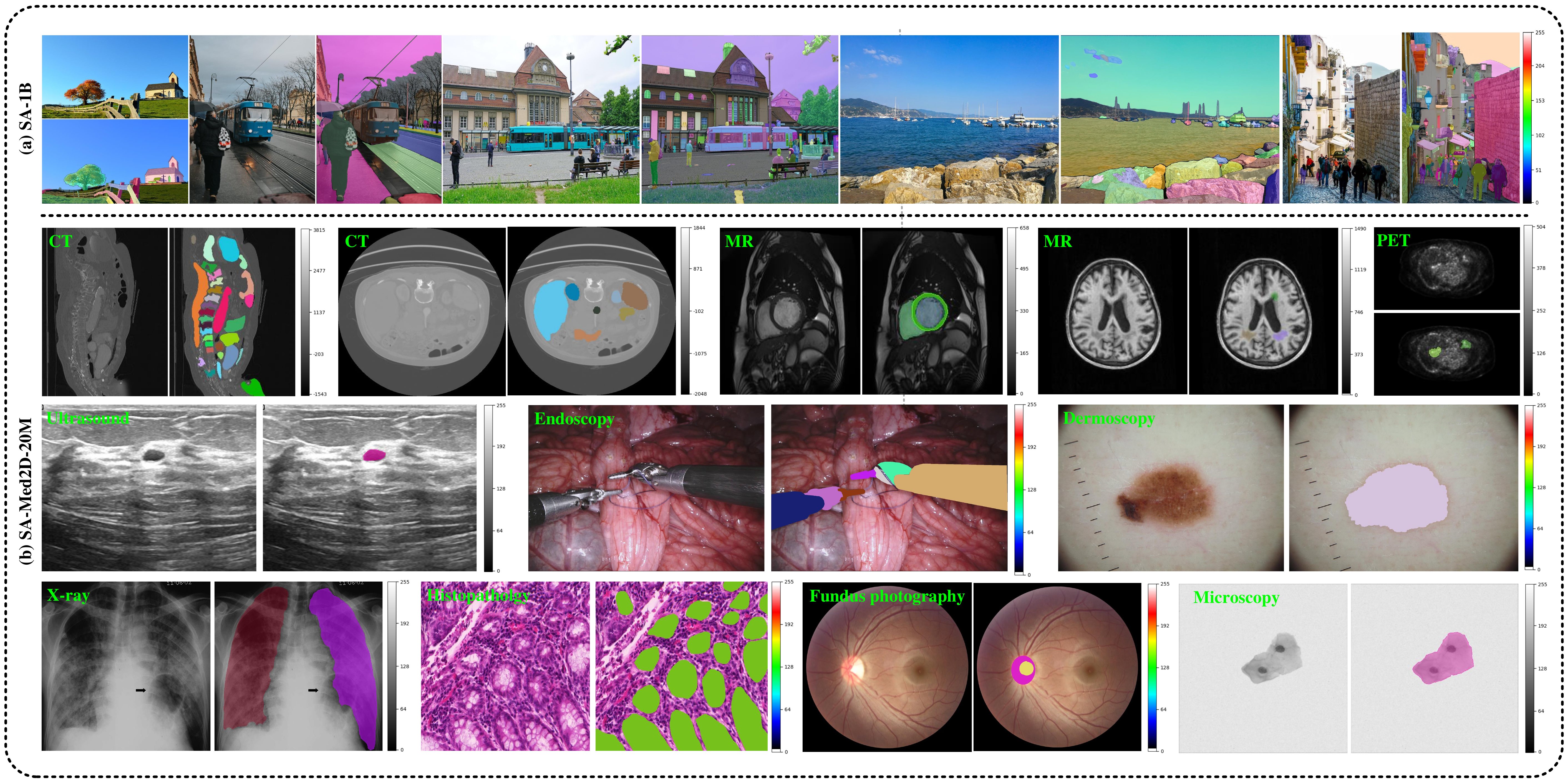}
    \caption{(a) Data format comparison. The data format of nature images is RGB with 3 channels and the pixel values range from 0 to 255, while medical images are various across different modalities.}
    \label{fig:intro_images}
  \end{subfigure}
  \begin{subfigure}[b]{1.0\textwidth}
    \centering
    \includegraphics[width=1.0\textwidth]{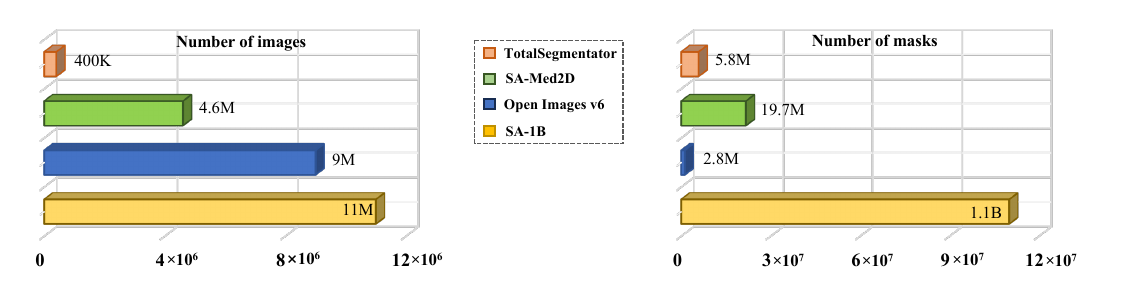}
    \caption{(b) Data size comparison. SA-1B consists of 11M natural images and corresponding 1129M masks, while the largest CT organ segmentation dataset TotalSegmentator only contains 400K valid CT slices and corresponding 5.8M masks. To bridge this gap, we formulate SA-Med2D-20M with 4.6M medical images and corresponding 19.7M masks.}
    \label{fig:intro_datasize}
  \end{subfigure}
\caption{Comparison of nature images and medical images.}
\label{fig:intro}
\end{figure}

Medical image segmentation plays a crucial role in diagnosing, radiotherapy planning, treating, and further medical research~\cite{huang2023stunet,isensee2021nnu,NEURIPS2022_ee604e1b}. Accurate segmentation for medical images is a challenging task as data modalities and targets can vary significantly in clinical practice. To deal with the complex scenarios in medical image segmentation, building medical vision foundation models is essential. Foundation models are usually trained on large-scale datasets and can excel in various data modalities and targets. As such, these models can help doctors in precisely identifying and locating areas of pathology in complex clinical scenarios, enabling more accurate diagnosis and treatment. The key to successful general-purpose foundation models is \textit{broad and diverse datasets}, which has been verified in natural image recognition (e.g., CLIP~\cite{radford2021learning}, SAM~\cite{kirillov2023segment}), nature language processing (NLP, e.g., LLaMA~\cite{touvron2023llama,touvron2023llama2}), and vision-language modeling (e.g., LLaVa~\cite{liu2023visual}, GPT4~\cite{openai2023gpt4}).
For instance, the Segment Anything Model (SAM) is trained on large datasets (1.1 billion masks) of which data are scraped from the Internet. However, due to the significant domain gap between natural images and medical ones, SAM cannot achieve satisfying segmentation results on multi-modal medical datasets. The huge domain gap can be attributed to the data collection methods: medical images are collected from specific protocols and scanners and are presented as different modalities (electrons, lasers, X-rays, ultrasound, nuclear physics, and magnetic resonance) due to their particular clinical purpose. As a result, the visual appearance of natural images and medical images is substantially different as reviewed in Figure~\ref{fig:intro_images}.


Building large-scale medical image datasets with annotated masks is the key to training a robust medical vision foundation model, which can accurately segment the target region that doctors or researchers are interested in. However, publicly available datasets for medical image segmentation are often limited in data size. Even the largest medical datasets are a fraction of the size and diversity of benchmark datasets in general computer vision~\cite{kirillov2023segment,wu2019tencent} and NLP~\cite{lehmann2015dbpedia,muhleisen2012web}. Figure~\ref{fig:intro_datasize} shows the data size gap between different research areas. More importantly, these public datasets often focus on a single environment, modality, or anatomical structure, thus with limited diversity. It is vital to build a large-scale dataset with diverse medical images, which will help to advance medical image analysis as successfully as computer vision and NLP.

\begin{figure}[h]
  \centering
  \includegraphics[width=1.0\textwidth]{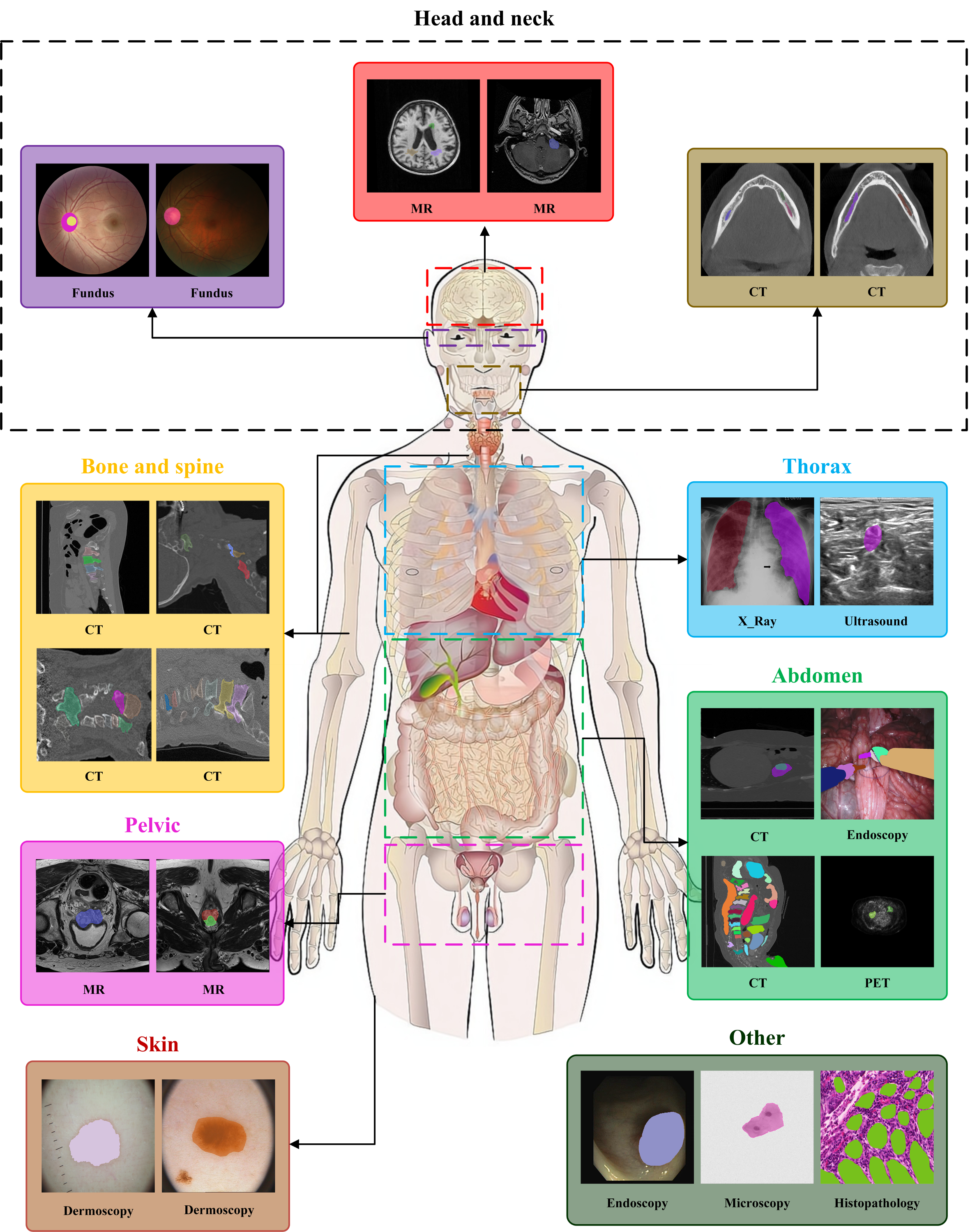}
  \caption{Overview of SA-Med2D-20M. It encompasses 4.6 million 2D medical images with 10 modalities and 19.7 million corresponding masks, representing a comprehensive coverage of nearly the whole body (over 200 categories).}
  \label{fig:sa_overview}
\end{figure}

Collecting a large-scale medical image dataset from scratch and annotating them carefully by specialists is infeasible~\cite{PARK202035} because accessing medical image data is not as easy as natural images and processing per-voxel annotations is expensive and time-consuming. To avoid the problems of data access and handcraft annotation, we introduce a new medical image dataset called ``SA-Med2D-20M'' which is built upon numerous existing datasets for the segmentation of molecules and cells to organ systems and the full body. 
Specifically, we take advantage of numerous public datasets on the web sources (TCIA~\footnote{\url{https://www.cancerimagingarchive.net}}, OpenNeuro~\footnote{\url{https://openneuro.org}}, NITRC~\footnote{\url{https://www.nitrc.org}}, Grand Challenge~\footnote{\url{https://grand-challenge.org}}, Synapse~\footnote{\url{https://www.synapse.org}}, CodaLab~\footnote{\url{https://codalab.lisn.upsaclay.fr}}, GitHub~\footnote{\url{https://github.com}}, etc) to collect as many public medical segmentation datasets as possible. Thanks to these resources,
SA-Med2D-20M is built to be a broad and diverse dataset, which consists of 4.6 million medical images and 19.7 million corresponding masks. To the best of our knowledge, it is the largest medical segmentation dataset by far. SA-Med2D-20M is also expected to encompass a wide range of modalities and categories. 
As shown in Figure~\ref{fig:sa_overview} (a summary of statistics of SA-Med2D-20M): the dataset contains 10 modalities, 31 main organs, and 271 labeled classes. This covers almost all object types in the currently available public datasets, aiming to address the deficiency of data in medical imaging. We hope that a large-scale dataset of medical segmentation is a helpful resource for developing advanced foundation models in medical image analysis.

The paper is organized as follows: we first provide an overview of SA-Med2D-20M in Section~\ref{sec:overview}, detailing what the dataset consists of and its properties from multiple perspectives. 
In section~\ref{sec:details}, we introduce the data collection and processing step by step. In section~\ref{sec:application}, we present one successful application by exploiting the current SA-Med2D-20M and discuss the potential values and limitations of the dataset.

\section{Related Works}
\subsection{Large-scale Vision Models (LVMs)}
LVMs~\cite{seggpt,sam,clip,dalle,seem,dosovitskiy2020image,jia2021scaling} usually have large-scale parameters which are trained on massive vision datasets. Owing to their impressive generalization abilities, they have demonstrated huge success in various tasks, such as image generation, classification, segmentation, etc. Among the LVMs, CLIP~\cite{clip} is renowned for bridging the vision and language domain by jointly pre-training the model on substantial image-text pairs. It can be applied to visual question answering, video understanding, and image manipulation. Similarly, DALL·E 2 ~\cite{dalle2} is trained on enormous data but targets generating images from text. Other notable LVMs include SegGPT~\cite{seggpt}, SEEM~\cite{seem}, and SAM~\cite{sam}, all trained on millions of images for segmentation tasks. Due to the generalizable knowledge learned from big data, they can also be extended for captioning, inpainting, and tracking tasks. 
These LVMs highlight the importance of large-scale datasets for general-purpose models, but there lack of large-scale medical image datasets for developing LVMs for medical image analysis. Without a large number of medical images for model development, these LVMs, usually trained on natural images only, cannot excel in medical tasks. This observation thus motivates us to collect the SA-Med2D-20M dataset for the medical domain.

\subsection{Medical Image Segmentation}
One of the most classical models for medical image segmentation is U-Net \cite{ronneberger2015u} and its variants, including U-Net++~\cite{zhou2018unet++}, ResU-Net~\cite{drozdzal2016importance}, nnU-Net~\cite{isensee2021nnu}. 
These methods are usually task-specific, which means that they can excel on a specific modality (e.g., CT) or target class (e.g., organs), but may fail on the others. The task-specific property may be a major limitation because modalities and target classes can vary significantly in clinical medical images. As such, the latest research focuses on general-purpose segmentation models for medical images. For instance, extensive studies~\cite{medlsam,sam3d,3dsamAdapt,masam,msa} have concentrated on applying SAM~\cite{sam} to medical image segmentation tasks. Since SAM~\cite{sam} is trained on natural images, it is necessary to incorporate medical knowledge into it so as to adapt it to the medical domain. To this end, some research resort to 1) fine-tuning the decoder of SAM \cite{17,20}, 2) designing adapters to the encoder of SAM \cite{21}, or 3) manipulating the prompt of SAM \cite{16,18}. These strategies are conducted on a limited number of medical images, which may affect their effectiveness. 
To further advance the development of medical image segmentation, large-scale medical datasets are crucial.

\subsection{Large datasets on medical imaging}
Collecting medical image datasets is labor-intensive and time-consuming. Thus, the image numbers of medical datasets are usually much less than those in natural image datasets. By far, some public large-scale medical image datasets, though containing multiple modalities and diverse organs/lesions, have only thousands of cases. For example, the AbdomenCT-1K \cite{ma2021abdomenct}, BraTS21 \cite{baid2021rsna}, AutoPET \cite{gatidis2022whole}, and TotalSegmentator \cite{wasserthal2022totalsegmentator} comprise over 1,000 annotated CT, MRI, or PET images, with the largest TotalSegmentator consisting of 1204 CT images of 104 annotated organs (1228 CT cases and 117 categories in ``v2'' version~\footnote{\url{https://github.com/wasserth/TotalSegmentator}}) for segmentation tasks. However, compared with millions of natural images used for LVM training, 1204 images can hardly support the development of advanced large-scale medical models. Our SA-Med2D-20M aims to fill this gap.

\section{Overview of SA-Med2D-20M}\label{sec:overview}
We curated SA-Med2D-20M by collecting medical images from public segmentation datasets on the web sources such as \href{https://www.cancerimagingarchive.net}{TCIA}, \href{https://openneuro.org}{OpenNeuro}, \href{https://www.nitrc.org}{NITRC}, \href{https://grand-challenge.org}{Grand Challenge}, \href{https://www.synapse.org}{Synapse}, \href{https://codalab.lisn.upsaclay.fr}{CodaLab}, \href{https://github.com}{GitHub}, etc. The SA-Med2D-20M encompasses 4.6 million 2D medical images and 19.7 million corresponding masks. In addition, it involves over 200 categories and 10 modalities, representing a comprehensive coverage of nearly the whole body and various imaging modalities. Figure~\ref{fig:sa_overview} provides a selection of exemplary images from the dataset.

\begin{figure}[t]
  \centering
  \begin{subfigure}[b]{0.48\textwidth}
    \centering
    \includegraphics[width=1.0\textwidth]{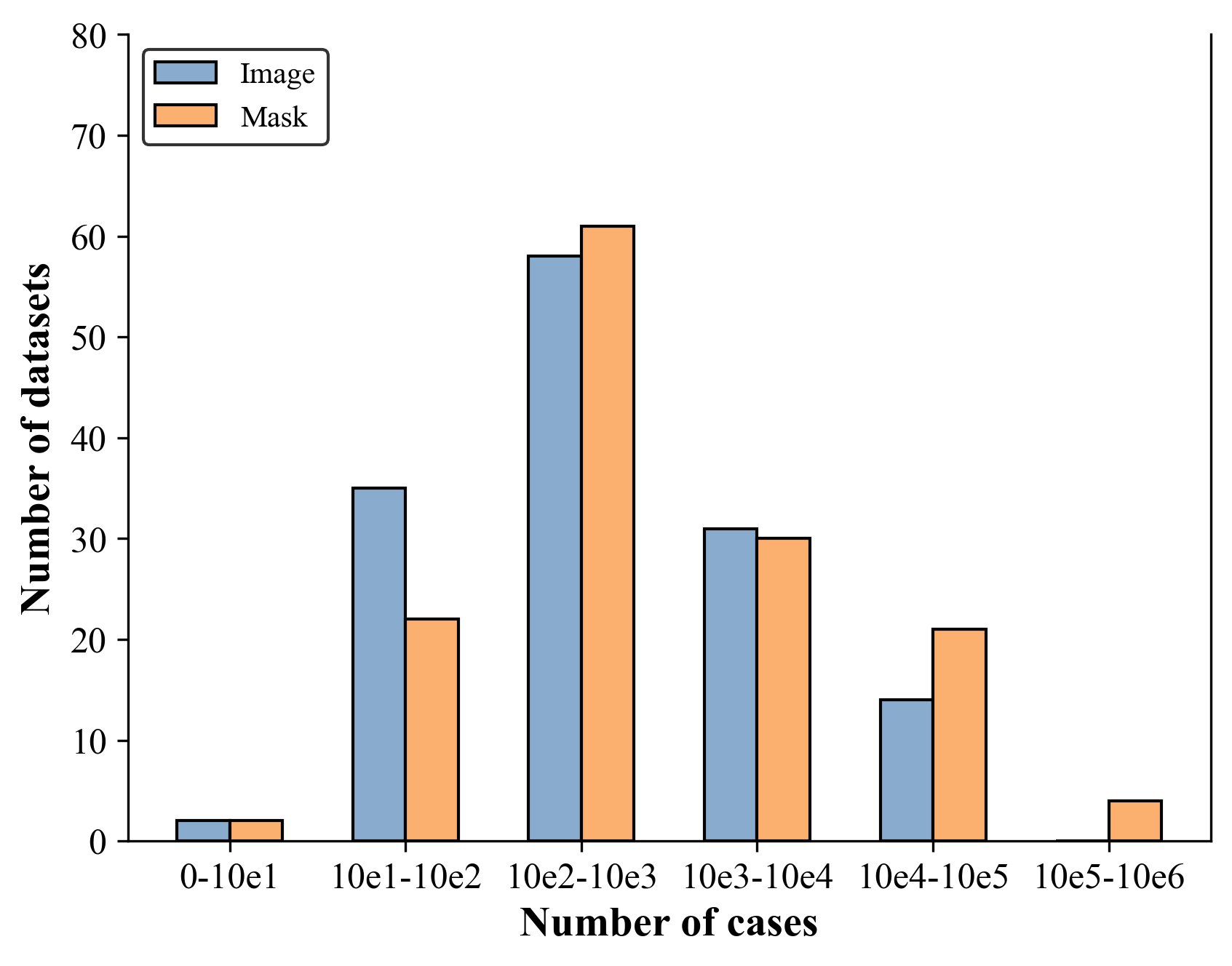}
    \caption{(a) The number of datasets categorized by the number of cases.}
    \label{fig:dataset_stat}
  \end{subfigure}
\hfill
  \begin{subfigure}[b]{0.48\textwidth}
    \centering
    \includegraphics[width=1.0\textwidth]{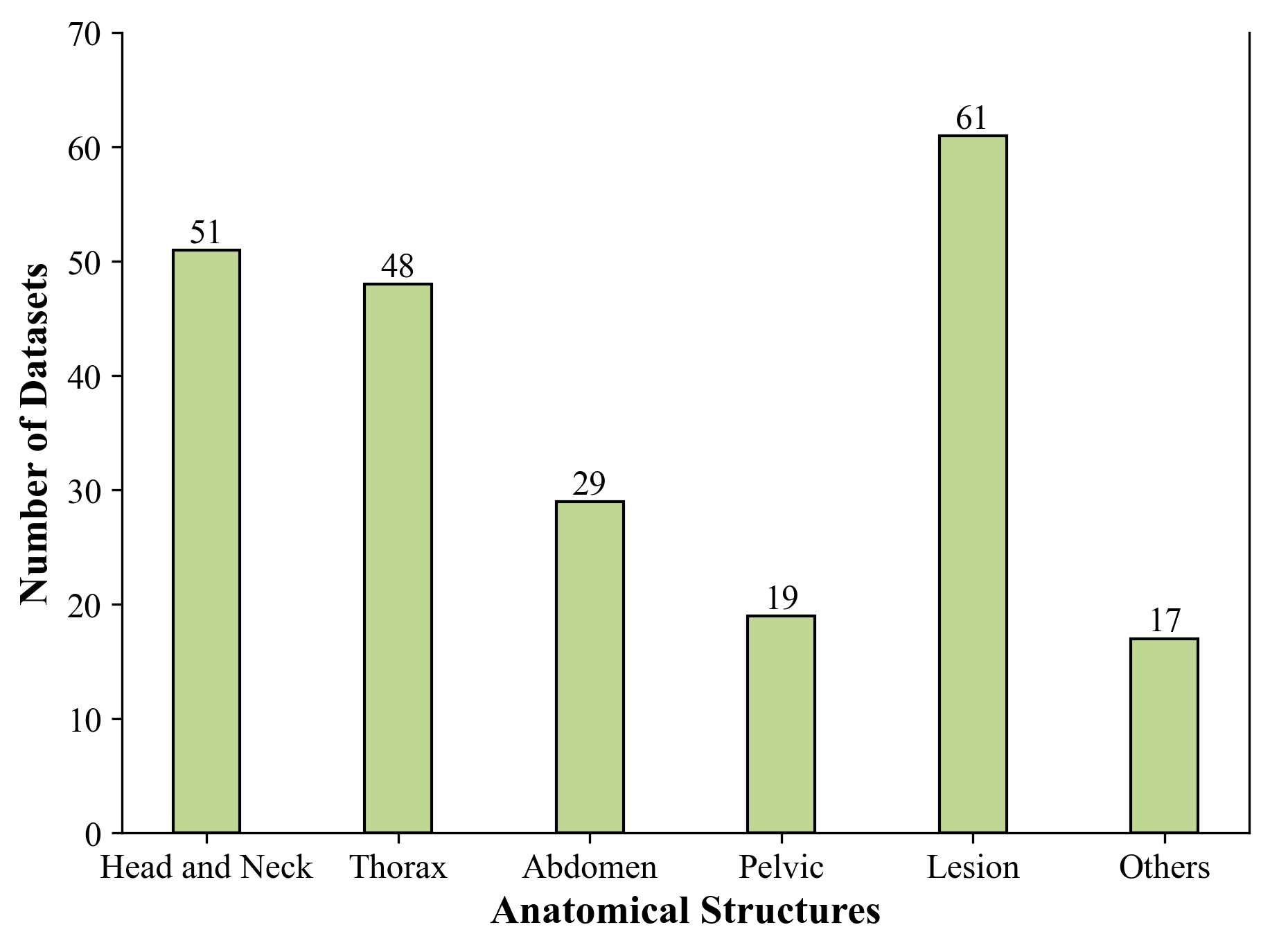}
    \caption{(b) The number of datasets categorized by anatomical structure.}
    \label{fig:anatomical_stats_dataset}
  \end{subfigure}
\caption{Distribution of datasets.}
\label{fig:stat_1}
\end{figure}

\textbf{Datasets.} The distribution of the datasets is depicted in Figure~\ref{fig:stat_1}. Despite the abundance of public medical image datasets available online, a significant proportion of them encompass only a limited number of cases. As illustrated in Figure~\ref{fig:dataset_stat}, over 120 datasets fall short of reaching 10,000 images or 100,000 masks. Furthermore, even the most extensive dataset in our collection contains fewer than 100,000 images and 1,000,000 masks, respectively. As a result, it is challenging to collect and release such a large and diverse medical image dataset. Figure~\ref{fig:anatomical_stats_dataset} presents the case distribution across various body parts in the SA-Med2D-20M dataset. The ``Lesion'' datasets constitute the largest proportion among all categories, suggesting a significant interest in automating lesion or tumor segmentation in medical images which is vital in assisting medical professionals. It is noteworthy that 17 datasets categorized under ``Other'' are difficult to assign to specific anatomical structures due to the lack of detailed anatomical information from the original dataset.

\begin{table}[htbp]
\centering
\begin{tabular}{c|c|c|c|c|c}
\hline
Modality & CT & Endoscopy & PET & Fundus & Microscopy \\
\hline
Images   & 2338753 & 5838 & 11956 & 3501 & 945 \\
Masks    & 12547037 & 19388 & 14129 & 7242 & 5894 \\
\hline\hline
Modality & MR & Dermoscopy & X-ray & Ultrasound & Histopathology \\
\hline
Images   & 2217633 & 7935 & 6407 & 2968 & 866 \\
Masks    & 7147784 & 8000 & 9357 & 2968 & 1056 \\
\hline
\end{tabular}
\caption{The number of images and masks of each modality. SA-Med2D-20M has 10 modalities in total. CT and MR predominate the number of both images and masks in the whole dataset.}
\label{tab:modal_stat}
\end{table}

\begin{figure}[t]
  \centering
  \begin{subfigure}[b]{0.48\textwidth}
    \centering
    \includegraphics[width=1.0\textwidth]{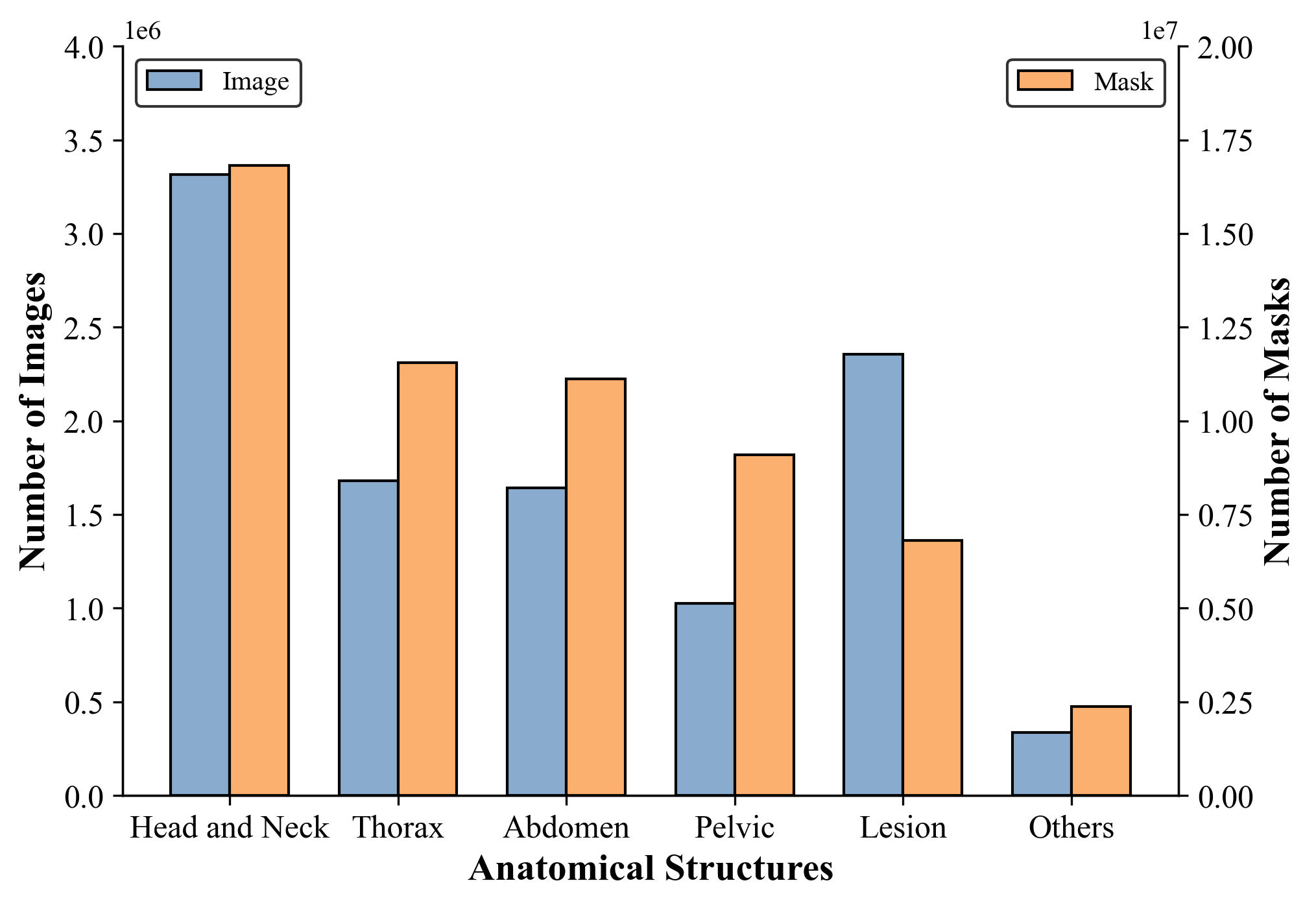}
    \caption{(a) The number of images and masks categorized by anatomical
structure.}
    \label{fig:anatomical_stats1}
  \end{subfigure}
\hfill
  \begin{subfigure}[b]{0.48\textwidth}
    \centering
    \includegraphics[width=1.0\textwidth]{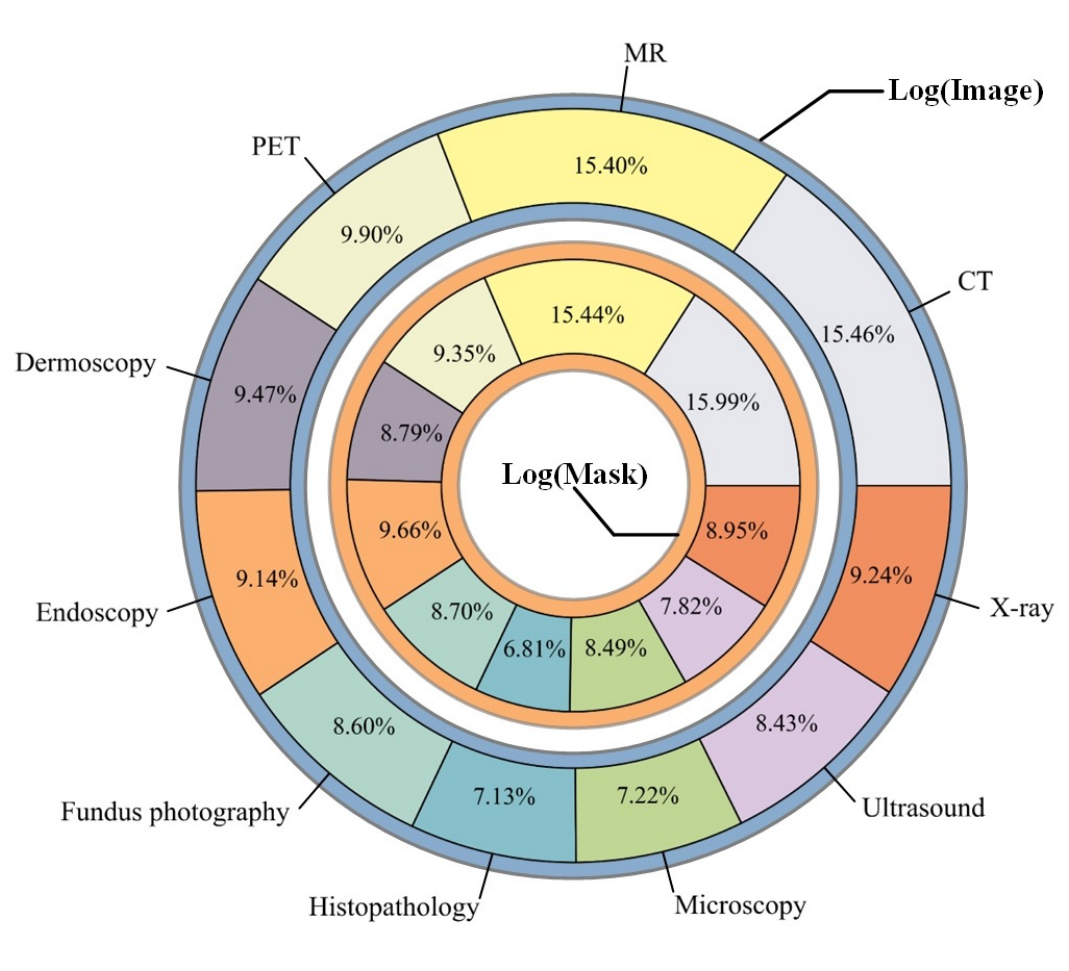}
    \caption{(b) The distribution of modalities (scaled logarithmically).}
    \label{fig:modal_stat}
  \end{subfigure}
\caption{Statistics of SA-Med2D-20M with anatomical structure and modalities, separately.}
\label{fig:stat_2}
\end{figure}

\textbf{Modality.} SA-Med2D-20M includes 10 modalities, as detailed in Table~\ref{tab:modal_stat}, with their distribution illustrated in Figure~\ref{fig:modal_stat}. CT and MR modalities are predominant in both the number of images and masks, it is mainly attributed to their widespread presence in public medical image segmentation datasets and the 3D dimension of CT and MR scans, which yields a high volume of slices when segmented across three axes. Following CT and MR, most modalities range between 1,000 to 10,000 images and 1,000 to 20,000 masks. Notably, Microscopy and Histopathology currently comprise fewer than 1,000 images each. We plan to add more data for these less-presented modalities in future updates of SA-Med2D-20M.

\textbf{Anatomical structures.} The dataset is classified into different categories based on anatomical structures and lesion presence, including \textit{Head and Neck}, \textit{Thorax}, \textit{Abdomen}, \textit{Pelvic}, \textit{Lesion}, and \textit{Others}, as illustrated in Figure~\ref{fig:anatomical_stats1}. The approximate ratio of the number of masks to images ranges from 3 to 10. The \textit{Head and Neck} category contains the largest number of both images and masks for the abundance of multi-modality brain-related data, such as from the BraTS and ISLES series. In contrast, the \textit{Others} category includes the fewest images and masks. As corroborated by Figure~\ref{fig:anatomical_stats_dataset}, tumor segmentation emerges as a primary focus in medical imaging. The \textit{Others} category refers to datasets not fitting into specific anatomical classes, such as those involving cell and skin segmentation.

\begin{figure}[ht!]
  \centering
  \begin{subfigure}[b]{0.32\textwidth}
    \centering
    \includegraphics[width=0.96\textwidth]{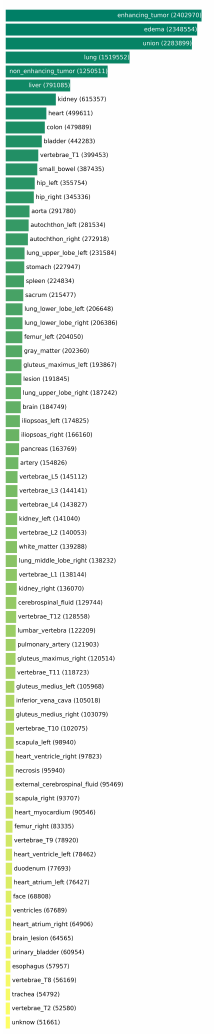}
    \label{fig:label_stat1}
  \end{subfigure}
\hfill
  \begin{subfigure}[b]{0.32\textwidth}
    \centering
    \includegraphics[width=0.96\textwidth]{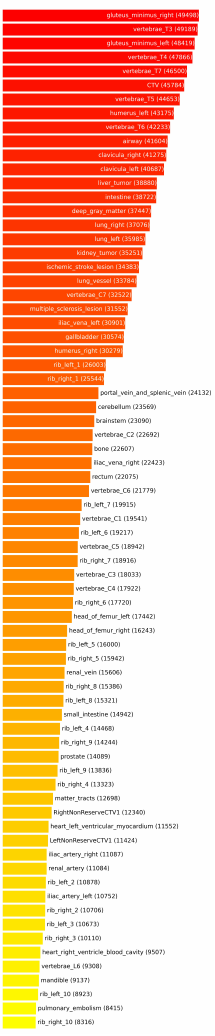}
    \label{fig:label_stat2}
  \end{subfigure}
\hfill
  \begin{subfigure}[b]{0.32\textwidth}
    \centering
    \includegraphics[width=0.96\textwidth]{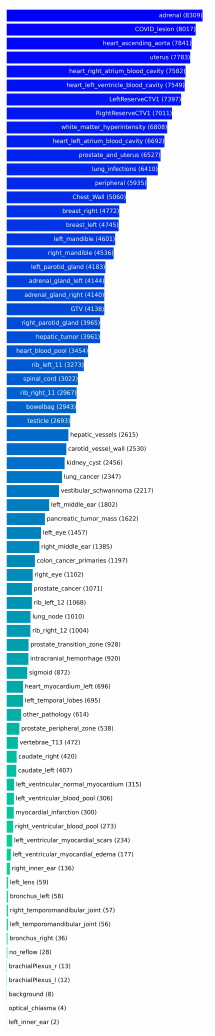}
    \label{fig:label_stat3}
  \end{subfigure}
\caption{Label distribution in SA-Med2D-20M. In total, the dataset comprises 219 original labels, with all label names are same as the released version. Note that the label ``unknown'' is used for instances where the original dataset has not provided specific label information.}
\label{fig:label_stat}
\end{figure}

\textbf{Categories.} SA-Med2D-20M consists of 219 labels, with a detailed distribution illustrated in Figure~\ref{fig:label_stat}. The categories ``enhancing\_tumor'' and ``edema'' from the BraTS datasets have the highest count. The ``union'' category is specifically designed to address the issue of pixel overlap across multiple classes whose masks cover more than two categories. Furthermore, the label ``unknown'' is assigned to instances where the original dataset does not provide specific label information. In terms of label distribution, the long-tailed problem exists in SA-Med2D-20M. Specifically, 47 categories fall within the range of 100,000 to 1,000,000 masks, and 51 categories between 1,000 and 10,000. The most common range is 10,000 to 100,000, with 88 categories lying in it. Additionally, there are also 28 categories with fewer than 1,000 masks.

\section{Construction of SA-Med2D-20M}\label{sec:details}
In this section, we describe how to collect and process cases to construct such a large-scale and diverse dataset in detail.

\subsection{Collecting data from public resources}
The first stage of the construction of SA-Med2D-20M is to collect candidate medical images. Different from general images that are abundant and easy to access on the Internet, we can only find medical images on specialized portal websites, such as TCIA, OpenNeuro, Grand Challenge, Synapse, GitHub, etc. After searching all medical-related open resources and selecting medical image segmentation datasets, we list detailed dataset information in Table~\ref{tab:all_datasets}. However, The raw datasets cannot be directly used for developing models because the dimension, modality, and intensity values of raw data are various. The next step is to normalize all datasets.

\begin{figure}[t]
  \centering
  \includegraphics[width=0.8\textwidth]{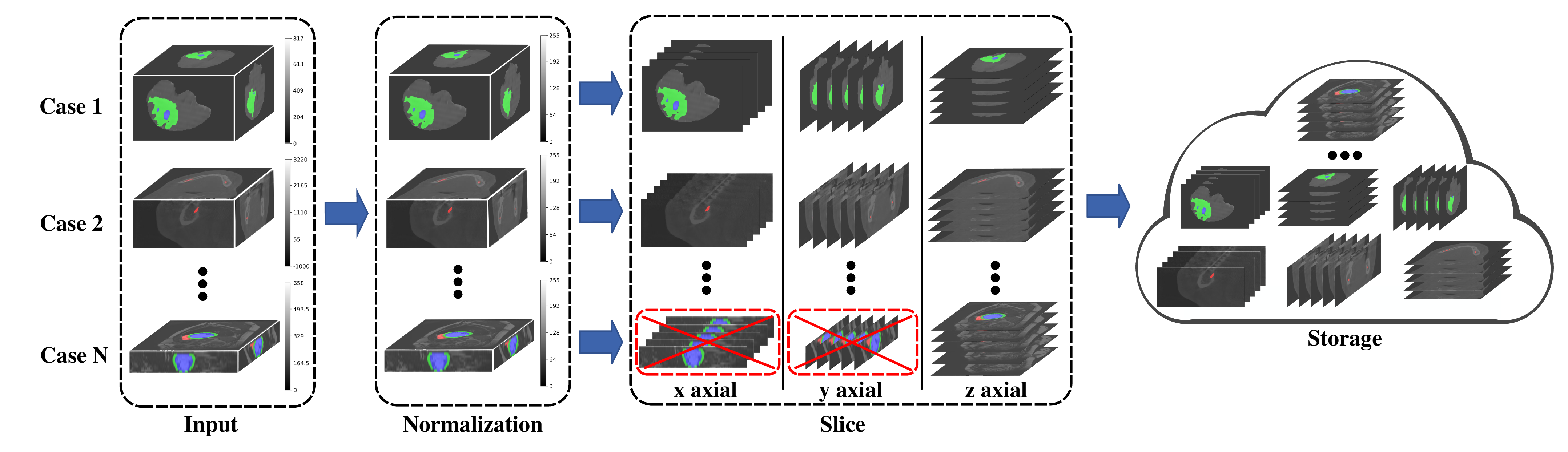}
  \caption{Pipeline of image normalization. Given an input image, we first normalize the voxel values to $[0, 255]$ with Self-Min-Max normalization. Then, we split them along three directions and discard images with extreme aspect ratios for 3D images. All processed images are saved in “PNG” format.}
  \label{fig:pipeline_s1}
\end{figure}
\subsection{Image normalization}
There are two aspects to be considered to normalize data. One is that the voxel or pixel values in medical images vary widely because of various modalities and collection methods. For example, the intensity range of CT is over 2,000, and MR is over 10,000. Another is how to unify the data dimension. Some modalities are 3D, like CT, MR, and PET, while some are 2D, like X-ray, Fundus, and Ultrasound. Figure~\ref{fig:pipeline_s1} illustrates the whole process of normalizing different datasets. In the first step, we normalize the voxel or pixel values to the \textit{``PNG''} format, which is detailed below. An original image is first normalized into $[0, 1]$ with the Self-Min-Max normalization (see Equation~\eqref{eq:normalize}), and then taken the ceiling value after multiplying the values by 255. The process can be formulated as:
\begin{equation}
\label{eq:normalize}
    x' = \lceil \frac{x - x_{min}}{x_{max}} \times 255 \rceil
\end{equation}
where $x_{min}$ and $x_{max}$ are the min value and max value in the image. In the second step, we split all 3D images along three axes and discard images with extreme aspect ratios. Specifically, slice images with the shortest edge less than half the length of the longest edge were discarded to prevent target areas from being extremely blurry. After that, all processed images are saved in \textit{``PNG''} format.

\begin{figure}[ht]
  \centering
  \includegraphics[width=0.8\textwidth]{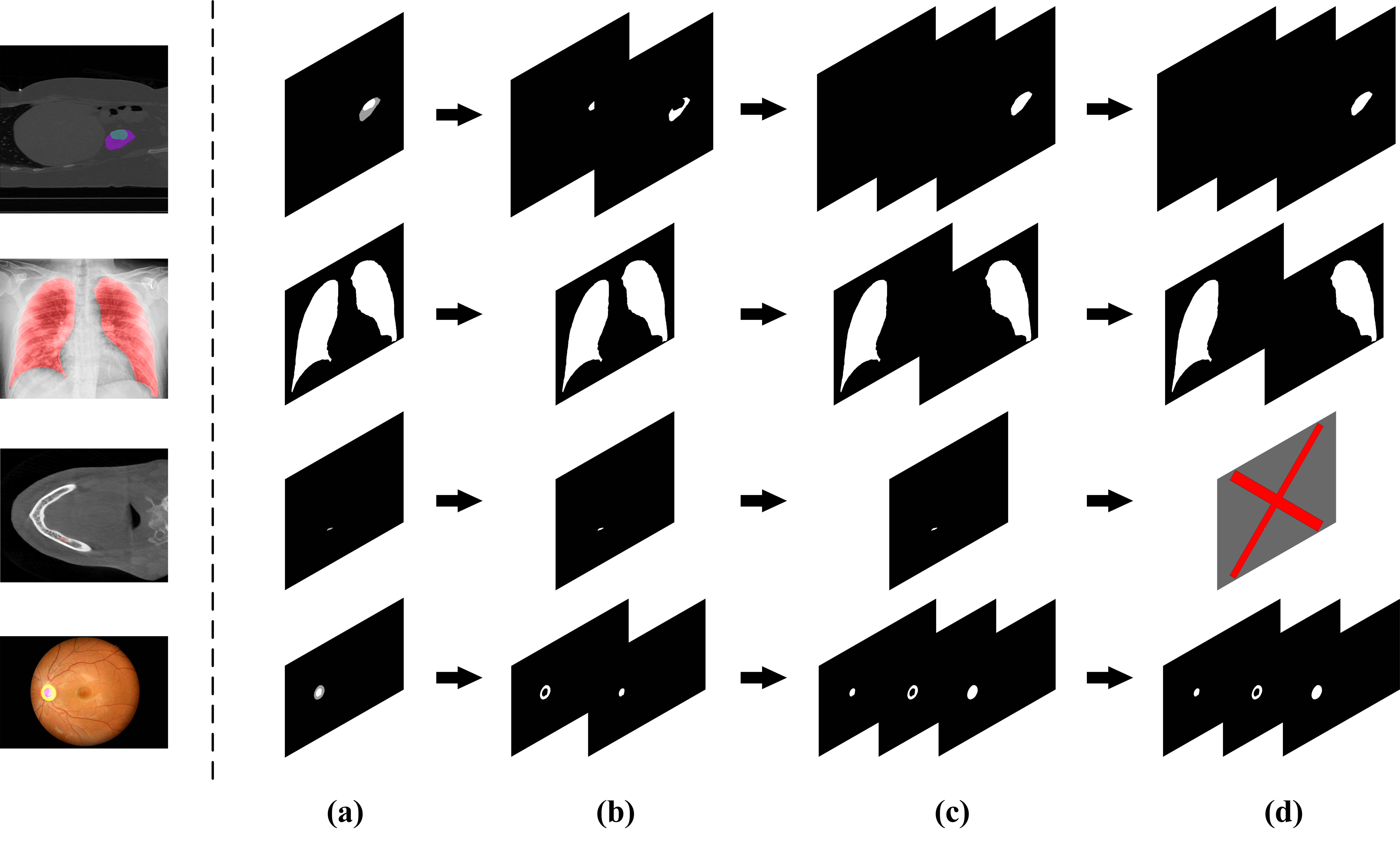}
  \caption{Overview of mask processing pipeline. (a)$\to$(b): original labels are split into binary masks. For example, if one mask contains two categories, we will generate two binary masks after this step. (b)$\to$(c): multiple connected components separation and overlapped areas consolidation. (c)$\to$(d): remove small masks.}
  \label{fig:pipeline_s2}
\end{figure}
\subsection{Mask processing}
We split multi-label masks into single foregrounds, the final representations are similar to SAM. Figure~\ref{fig:pipeline_s2} shows the process flow. It can be summarized as three steps. (1) split semantic mask to binary mask. (2) separate binary foreground to multiple connected components and consolidate overlapped areas. (3) remove masks that do not meet our criteria. Then we will describe them in detail. 


\begin{figure}[t]
  \centering
  \begin{subfigure}[b]{0.9\textwidth}
    \centering
    \includegraphics[width=0.9\textwidth]{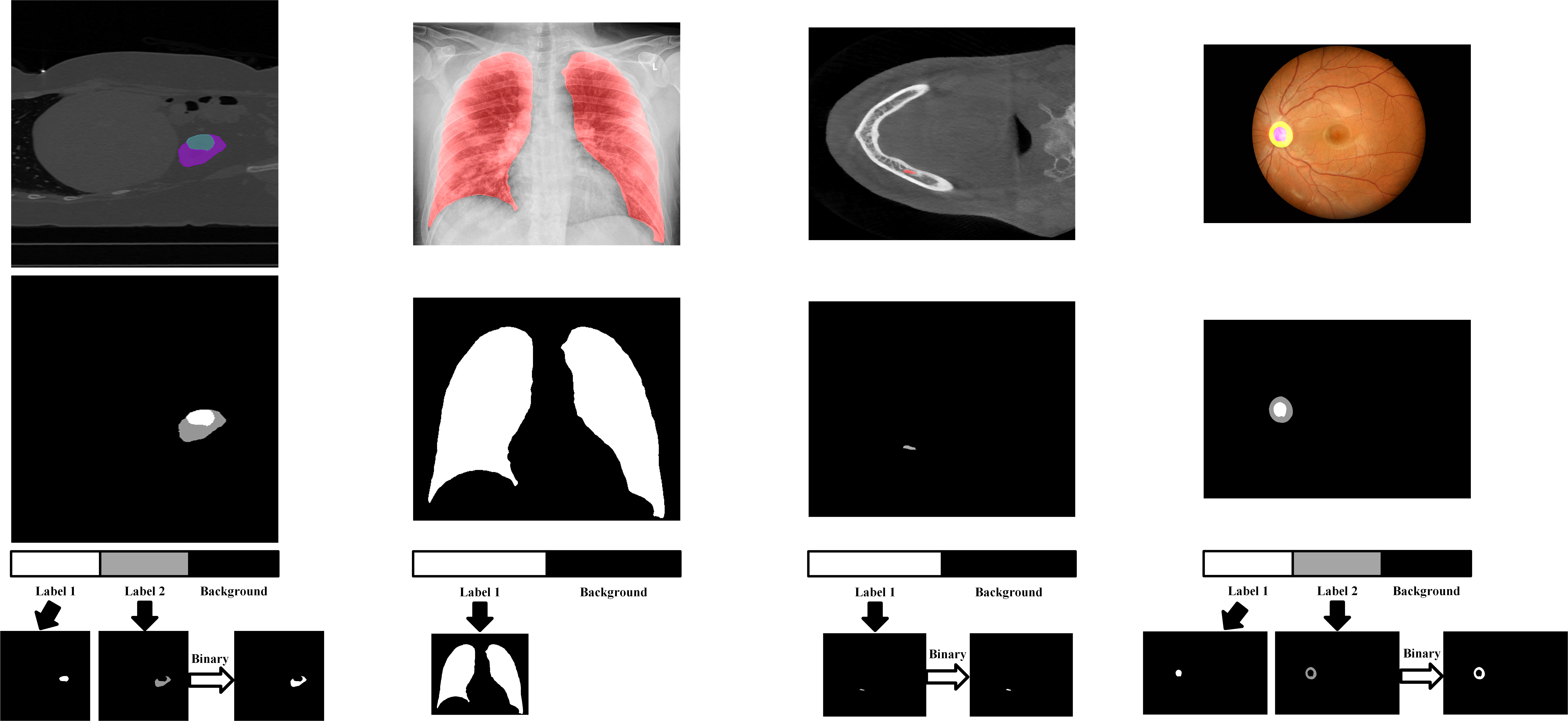}
    \caption{First Step: four examples that we split the original masks into binary masks.}
    \label{fig:detail_s21}
  \end{subfigure}
\hfill
  \begin{subfigure}[b]{0.9\textwidth}
    \centering
    \includegraphics[width=0.9\textwidth]{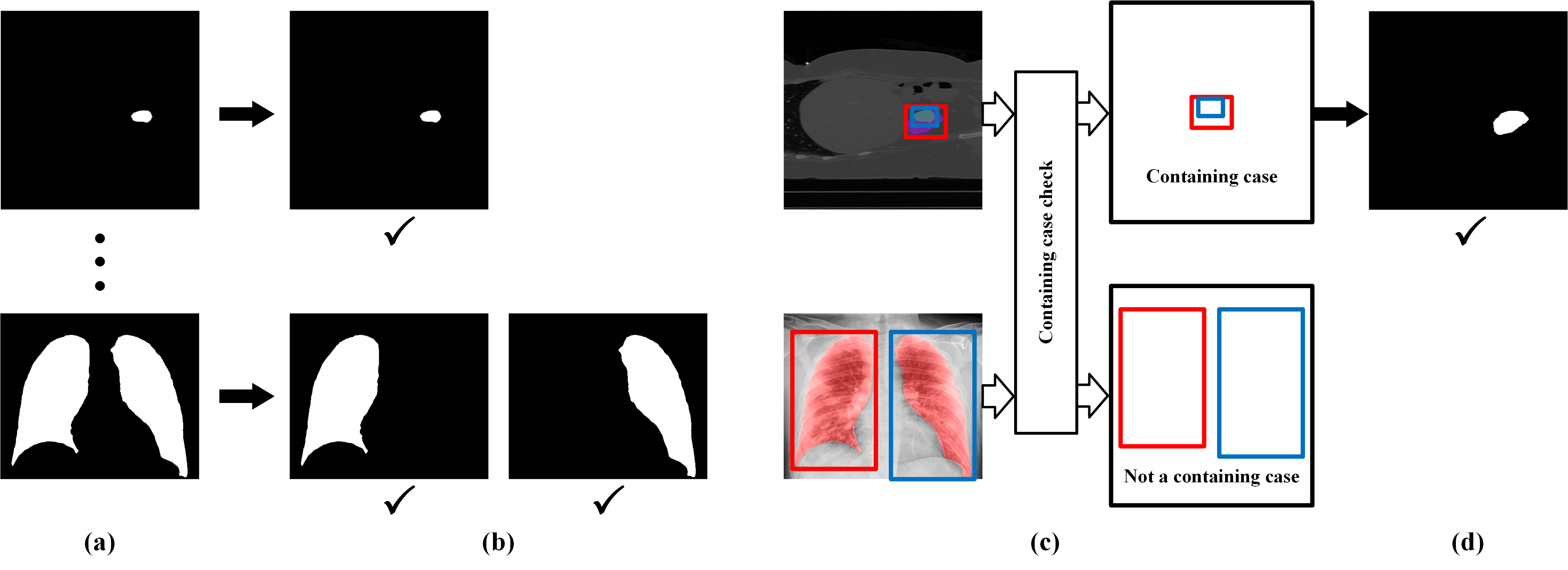}
    \caption{Second step: separation ((a)$\to$(b)) and consolidation ((c)$\to$(d))operations.}
    \label{fig:detail_s22}
  \end{subfigure}
\hfill
  \begin{subfigure}[b]{0.9\textwidth}
    \centering
    \includegraphics[width=0.9\textwidth]{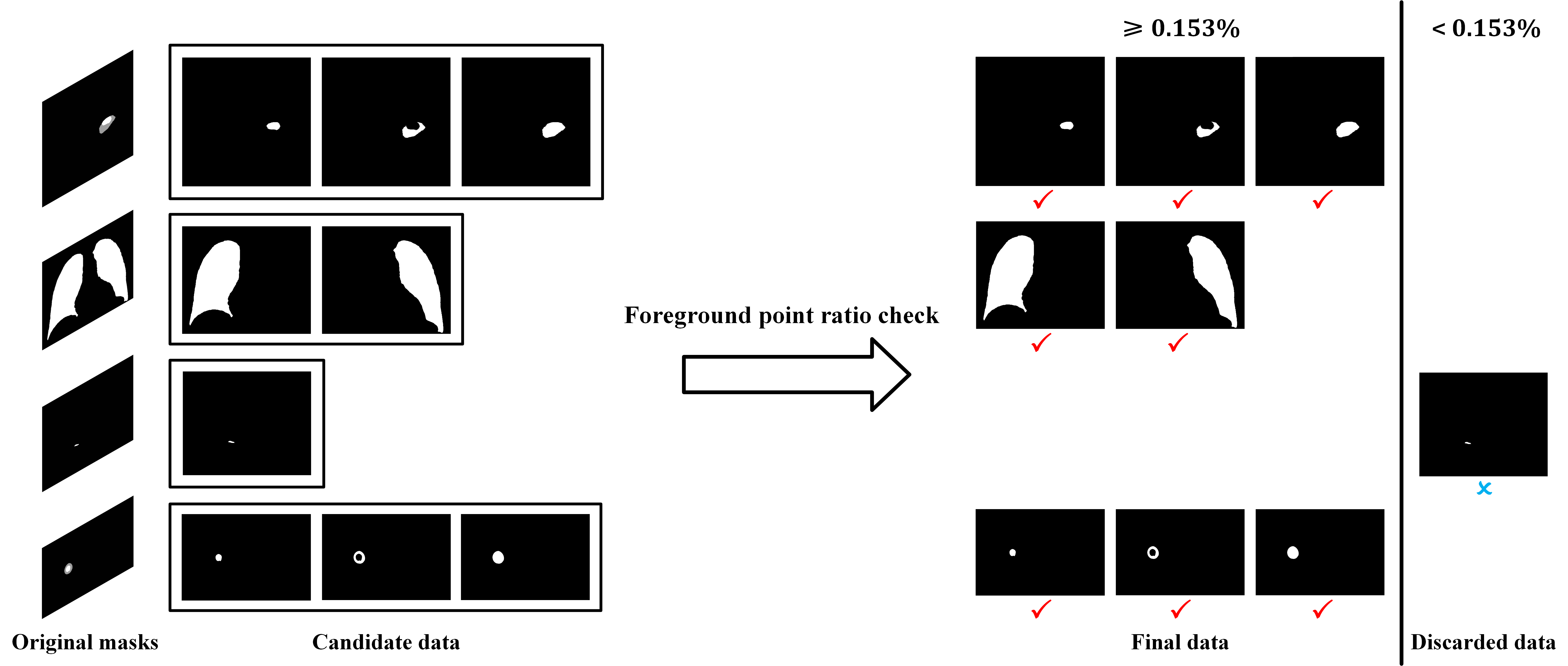}
    \caption{Third step: The processed masks that fulfill the area criteria will be retained.}
    \label{fig:detail_s23}
  \end{subfigure}
\caption{Details of mask processing.}
\label{fig:detail_s2}
\end{figure}

The first step is to split original labels into binary foregrounds. As shown in Figure~\ref{fig:detail_s21}, given an original mask image, we extract the foreground images according to label IDs. Then all separated foregrounds are assigned the value 1 while backgrounds are assigned 0. 
The second step involves two operations as two examples illustrated in Figure~\ref{fig:detail_s22}: one operation is the separation of foregrounds into distinct connected components, each representing a unique category such as individual cells in pathology images or discrete organs in CT scans. The other operation is to consolidate smaller segmentation targets (e.g., a tumor) contained in a larger target (e.g., a kidney containing that tumor) so that the overlap area of different classes can have only a single label (e.g., a kidney). 
For instance, the ((c)$\to$(d)) process in the first case shown in Figure~\ref{fig:detail_s22} illustrates how to consolidate foreground elements into a larger entity when a single kidney contains tumors.
In the third step as shown in Figure~\ref{fig:detail_s23}, masks are filtered out if the target area accounts for less than 0.153\% ($\frac{100}{256{\times}256}$) of the total image area -- equivalent to 100 pixels in a $256{\times}256$ resolution. This criterion ensures that only regions of sufficient size are considered, enhancing the precision and decreasing the noise of our analysis.

After the whole processing, each image is associated with multiple masks, and the original category information for each mask is discarded. To address this, we create a \textit{``JSON''} file to record the category information for each mask systematically. Then, we establish a uniform naming convention for images and masks, separately. The name format can be formulated as:
\begin{small}
\begin{align}
    - & Images: \\
      & -\{modality\}--\{dataset\}--\{ori\_name\}--\{dim\_slice\_id\}.png \\
    - & Mask:  \\
      & -\{modality\}--\{dataset\}--\{ori\_name\}--\{dim\_slice\_id\}--\{instance\_id\}.png
\end{align}
\end{small}
where $dataset$ refers to the specific dataset name that the case is from. $ori\_name$ is the original case name in its dataset. $dim\_slice\_id$ is mainly for 3D images: If we split a 3D case with axis x and the current slice is 100, then the term can be ``$x\_0100$''. Besides, we assign ``$2d\_none$'' to all 2D cases in this field. $instance\_id$ is unique to masks and encapsulates both category information and instance id, and the detailed information is stored in \textit{``JSON''} file. For instance, if the category ``liver'' is assigned the ID ``0003'' and there is only one instance of this category in the case, the $instance\_id$ can be denoted as ``$0003\_000$'' and the category ``liver'' in \textit{``JSON''} file is formulated as key-value pair with \textit{python-dict} format: \{``liver'': ``0003''\}. Finally, all validated masks are saved in the \textit{``PNG''} format for consistency and ease of use.


\section{Applications and Discussions}\label{sec:application}
\textbf{Medical Vision Foundation Models.}
The vision foundation model has evolved from the pre-training and fine-tuning paradigm which is a model that is ``trained on broad data at scale and is adaptable to a wide range of downstream tasks''~\cite{bommasani2021opportunities}. On the one hand, though vision foundation models in natural imaging like SAM~\cite{kirillov2023segment} and CLIP~\cite{radford2021learning} have had impressive performance in various vision fields~\cite{ji2023sam,tang2023sam}, they cannot achieve satisfying performance on medical imaging for the significant domain gap between nature images and medical ones. On the other hand, the relative scarcity of medical imaging data hinders the development of medical foundation models. Our work, however, provides the possibility to bridge the gap between natural and medical imaging by providing a large-scale medical image dataset, which can serve as a training resource for both supervised training and self-supervised learning. For example, self-supervised techniques (MoCo~\cite{he2020momentum}, MAE~\cite{he2022masked}) are readily utilized on 4.6 million 2D medical images in our dataset, and training a medical SAM becomes feasible with 19.7 million masks.

\textbf{Exploiting SA-Med2D-20M.}
While transfer learning techniques on pre-trained models are useful for adapting knowledge to downstream tasks with limited data, directly training models from scratch using massive task-specific datasets is more popular and effective in medical image analysis. SA-Med2D-20M is well-organized by modality, anatomical structure, and label, which allows researchers to efficiently select data pertinent to their specific applications, significantly reducing the time required for data collection. Finally, we believe it can significantly alleviate the challenge of data acquisition and potentially accelerate the process of data iteration in medical image analysis after SA-Med2D-20M has been released.

\textbf{Limitations and Future work.}
Despite the vast number of images and masks in SA-Med2D-20M, the initial version of the dataset meets several limitations. The first limitation is mask integrity: SA-Med2D-20M may lack masks for some segmentation targets. Since we rely on the original labels for each image in the preliminary release which only provides annotated masks of interest for a specific task, the remaining ``regions of non-interest'' do not have masks but can be the segmentation targets in SA-Med2D-20M. In this case, we may miss the masks for these regions (i.e., regions of non-interest for such a specific task but of interest in our label system). 
The second one is data imbalance: SA-Med2D-20M currently faces a marked long-tailed problem due to data imbalance, especially in certain modalities and categories where public segmentation datasets are limited. The third is data format unification: our current approach is to simply convert the original data into RGB format using Self-Min-Max normalization. However, this approach cannot fully preserve the data information, and finding an effective way to normalize data formats across diverse modalities remains a significant and ongoing challenge.

In future work, we will try to pursue this work with the following aspects. For mask integrity, we intend to adopt a methodology similar to the data engine in SAM~\cite{kirillov2023segment}, which tags all data with pseudo labels and refines them manually. For data imbalance, 
we prepare to enrich SA-Med2D-20M by collecting additional data where there are current shortages in the initial version and applying the pseudo-labeling strategy, used to address mask integrity issue, on publicly available unlabeled data from open sources. 
Moreover, we encourage the community to contribute towards building a more comprehensive medical image segmentation dataset.  

\section{Acknowledgements}
We thank all medical workers and dataset owners for making public datasets available to the community.

\section*{\centering Appendix}
\begin{center}
\small
\begin{longtable}{lllclc}
\hline
\textbf{ID} & \textbf{Dataset Name}    & \textbf{Modality}   & \textbf{Dim} & \textbf{Anatomical structures}       & \textbf{Release} \\ \hline
\endfirsthead
\multicolumn{5}{c}{{\tablename\ \thetable{} : Continued from previous page}} \\ \hline
\textbf{ID} & \textbf{Dataset Name}    & \textbf{Modality}   & \textbf{Dim} & \textbf{Anatomical structures}       & \textbf{Release} \\ \hline
\endhead
1 & ACDC~\cite{19}                                                                                            & MR                 & 3D        & Thorax                               & \textcolor{green}{\ding{52}}        \\
2 & AMOS22~\cite{ji2022amos}                                                                                       & CT, MR              & 3D        & Thorax, Abdomen, Pelvic                & \textcolor{green}{\ding{52}}        \\
3 & ATM22~\cite{zhang2023multi}                                                                                    & CT                 & 3D        & Thorax                               & \textcolor{green}{\ding{52}}        \\
4 & AbdomenomenCT-1K~\cite{ma2021abdomenct}                                                                            & CT                 & 3D        & Abdomen                              & \textcolor{green}{\ding{52}}        \\
5 & ASC18~\cite{xiong2021global}                                                 & MR                 & 3D        & Thorax                               & \textcolor{green}{\ding{52}}        \\
6 & COSMOS22~\cite{chen2022carotid}                                                                              & MR                 & 3D        & H\&N                       & \textcolor{green}{\ding{52}}        \\
7 & BTCV~\cite{landman2015miccai}                                                                                  & CT                 & 3D        & Thorax, Abdomen, Pelvic                & \textcolor{green}{\ding{52}}        \\
8 & BraTS2013~\cite{menze2014multimodal, info:doi/10.2196/jmir.2930}                                              & MR                 & 3D        & H\&N, $\star$                & \textcolor{green}{\ding{52}}        \\
9 & BraTS2015~\cite{menze2014multimodal, info:doi/10.2196/jmir.2930}                                              & MR                 & 3D        & H\&N, $\star$                & \textcolor{green}{\ding{52}}        \\
10 & BraTS2018~\cite{menze2014multimodal, bakas2017advancing, bakas2019identifying}                                & MR                 & 3D        & H\&N, $\star$                & \textcolor{green}{\ding{52}}        \\
11 & BraTS2019~\cite{menze2014multimodal, bakas2017advancing, bakas2019identifying}                                & MR                 & 3D        & H\&N, $\star$                & \textcolor{green}{\ding{52}}        \\
12 & BraTS2020~\cite{menze2014multimodal, bakas2017advancing, bakas2019identifying}                                & MR                 & 3D        & H\&N, $\star$                & \textcolor{green}{\ding{52}}        \\
13 & BraTS2021~\cite{bakas2017advancing, bakas2019identifying, baid2021rsnaasnrmiccai}                             & MR                 & 3D        & H\&N, $\star$                & \textcolor{green}{\ding{52}}        \\
14 & BrainPTM2021~\cite{avital2019neural,nelkenbaum2020automatic}                                                  & MR                 & 3D        & H\&N, $\star$                & \textcolor{green}{\ding{52}}        \\
15 & CAD-PE~\cite{gonzález2020computer}                                                                                           & CT                 & 3D        & Thorax, $\star$                       & \textcolor{green}{\ding{52}}        \\
16 & CAUSE07~\cite{van20073d}                                                                                       & MR                 & 3D        & H\&N                       & -     \\
17 & CHAOS~\cite{CHAOS2021,CHAOSdata2019,kavur2019}                                                                 & MR                 & 3D        & Abdomen                              & \textcolor{green}{\ding{52}}        \\
18 & CMRxMotion~\cite{wang2022extreme}                                                                              & MR                 & 3D        & Thorax                               & \textcolor{green}{\ding{52}}        \\
19 & COVID-19 CT scans~\cite{paiva2020helping, jun2020covid}                                                        & CT                 & 3D        & Thorax, $\star$                       & \textcolor{green}{\ding{52}}        \\
20 & COVID-19-20~\cite{roth2022rapid}                                                                               & CT                 & 3D        & Thorax, $\star$                              & \textcolor{green}{\ding{52}}        \\
21 & COVID-19-Image~\cite{cohen2020covid, cohen2020covidProspective}                                                & X-ray              & 2D        & Thorax, $\star$                              & \textcolor{green}{\ding{52}}        \\ 
22 & CRASS~\cite{hogeweg2012clavicle}                                                                               & X-ray              & 2D        & Thorax                               & \textcolor{green}{\ding{52}}        \\ 
23 & CTPelvic1K~\cite{liu2021deep}                                                                                  & CT                 & 3D        & Pelvic                               & \textcolor{green}{\ding{52}}        \\
24 & CTSpine1K~\cite{deng2021ctspine1k}                                                                             & CT                 & 3D        & H\&N, Thorax, Abdomen         & \textcolor{green}{\ding{52}}        \\
25 & CVC-ClinicDB~\cite{bernal2015wm}                                                                               & EGD          & 2D        & Other, $\star$                              & \textcolor{green}{\ding{52}}        \\
26 & MosMedData COVID19~\cite{soham1024}                               & CT                 & 3D        & Thorax, $\star$                              & \textcolor{red}{\ding{56}}       \\
27 & Chestimage~\cite{tianchi83075}                    & X-ray              & 2D        & Thorax, $\star$                              & \textcolor{green}{\ding{52}}        \\
28 & Cranium~\cite{hssayeni2020computed}                       & CT                 & 2D        & H\&N                       & \textcolor{green}{\ding{52}}        \\
29 & CrossMoDA2021~\cite{dorent2023crossmoda}                                                                       & MR                 & 3D        & H\&N, $\star$                & \textcolor{green}{\ding{52}}        \\
30 & CrossMoDA2022~\cite{shusharina2021cross}                                                                       & MR                 & 3D        & H\&N, $\star$                & \textcolor{green}{\ding{52}}        \\
31 & DRISHTI-GS~\cite{sivaswamy2015comprehensive, sivaswamy2014drishti}                                             & FP & 2D        & H\&N                       & -     \\
32 & EMIDEC~\cite{lalande2022deep}                                                                                  & MR                 & 3D        & Thorax, $\star$                       & \textcolor{green}{\ding{52}}        \\
33 & EndoVis15~\cite{bernal2017comparative}                                                                         & EGD          & 2D        & Other, $\star$                              & \textcolor{green}{\ding{52}}        \\
34 & EndoCV2020~\cite{ali2020endoscopy}                                                              & EGD          & 2D        & Other, $\star$                              & -     \\
35 & FLARE21~\cite{MedIA-FLARE21}                                                                                   & CT                 & 3D        & Abdomen                              & \textcolor{green}{\ding{52}}        \\
36 & FLARE22~\cite{FLARE22}                                                                                         & CT                 & 3D        & Thorax, Abdomen                       & \textcolor{green}{\ding{52}}        \\
37 & FeTA2021~\cite{payette2021automatic}                                                                          & MR                 & 3D        & H\&N, Thorax                 & \textcolor{red}{\ding{56}}       \\
38 & FeTA2022~\cite{payette2021automatic}                                                                          & MR                 & 3D        & H\&N, Thorax                 & \textcolor{red}{\ding{56}}       \\
39 & Fusc2021~\cite{wang2022fuseg}                                         & DS     & 2D        & Other, $\star$                              & \textcolor{green}{\ding{52}}        \\
40 & GLaS~\cite{sirinukunwattana2015stochastic, sirinukunwattana2017gland}                                         & HP     & 2D        & Other, $\star$                              & \textcolor{red}{\ding{56}}        \\
41 & HVSMR2016~\cite{pace2015interactive}                                                                          & MR                 & 3D        & Thorax                               & \textcolor{green}{\ding{52}}        \\
42 & Heart Seg MRI~\cite{tobon2015benchmark}                                   & MR                 & 3D        & Thorax                               & \textcolor{green}{\ding{52}}        \\
43 & ADAM~\cite{fu2020adam}                                                                            & FP & 2D        & H\&N                       & \textcolor{green}{\ding{52}}        \\
44 & REFUGE2~\cite{orlando2020refuge, li2020development}                                               & FP & 2D        & H\&N                       & -     \\
45 & PALM19~\cite{huazhu2019palm}                                                                      & FP & 2D        & H\&N                       & \textcolor{green}{\ding{52}}        \\
46 & GAMMA~\cite{8252743,orlando2020refuge,fu2020age}                                                  & FP & 2D        & H\&N                       & \textcolor{green}{\ding{52}}        \\
47 & ISLES2015-~\cite{maier2017isles}                                                                               & MR                 & 3D        & H\&N, $\star$                              & \textcolor{green}{\ding{52}}        \\
48 & ISLES2016~\cite{winzeck2018isles}                                                                             & MR                 & 3D        & H\&N, $\star$                              & \textcolor{green}{\ding{52}}        \\
49 & ISLES2017~\cite{winzeck2018isles}                                                                             & MR                 & 3D        & H\&N, $\star$                              & \textcolor{green}{\ding{52}}        \\
50 & ISLES2018~\cite{cereda2016benchmarking, hakim2021predicting}                                                  & CT                 & 3D        & H\&N, $\star$                              & \textcolor{green}{\ding{52}}        \\
51 & ISLES2022~\cite{hernandez2022isles}                                                                           & MR                 & 3D        & H\&N, $\star$                              & \textcolor{green}{\ding{52}}        \\
52 & InSTANCE2022~\cite{li2023stateoftheart, 9511297}                                                               & CT                 & 3D        & H\&N, $\star$                & \textcolor{green}{\ding{52}}        \\
53 & KiPA22~\cite{HE2021102055, HE2020101722, SHAO2011849, SHAO20121001}                                            & CT                 & 3D        & Abdomen, $\star$                      & -     \\
54 & KiTS19~\cite{heller2019kits19}                                                                                 & CT                 & 3D        & Abdomen, $\star$                      & \textcolor{green}{\ding{52}}        \\
55 & KiTS21~\cite{zhao2021coarse}                                                                                   & CT                 & 3D        & Abdomen, $\star$                      & \textcolor{green}{\ding{52}}        \\
56 & LAScarQS2022~\cite{LI2022102303, LI2022102360, li2021atrialgeneral}                                           & MR                 & 3D        & Thorax, $\star$                       & -     \\
57 & LNDb~\cite{pedrosa2019lndb}                                                                                    & CT                 & 3D        & Thorax, $\star$                              & \textcolor{green}{\ding{52}}        \\
58 & LUNA16~\cite{setio2017validation}                                                                              & CT                 & 3D        & Thorax                               & \textcolor{green}{\ding{52}}        \\
59 & STACOM SLAWT~\cite{karim2018algorithms}                                & MR                 & 3D        & Thorax                               & -     \\
60 & LiTS~\cite{bilic2019liver}                                                                                     & CT                 & 3D        & Abdomen, $\star$                      & -     \\
61 & LMSLS~\cite{CARASS201777}                                        & MR                 & 3D        & Other, $\star$                              & \textcolor{green}{\ding{52}}        \\
62 & M\&Ms-2~\cite{campello2021multi}                                                                                     & MR                 & 3D        & Thorax                               & \textcolor{green}{\ding{52}}        \\
63 & MM-WHS~\cite{zhuang2018multivariate, zhuang2016multi, luo2022mathcal} & CT, MR              & 3D        & Thorax                               & \textcolor{green}{\ding{52}}     \\
64 & MRBrains13~\cite{mendrik2015mrbrains}                                                                          & MR                 & 3D        & H\&N                       & -     \\
65 & NEATBrainS15~\cite{mendrik2015mrbrains}                                                                        & MR                 & 3D        & H\&N                       & -     \\
66 & MRBrains18~\cite{mrbrains18}                                                                                   & MR                 & 3D        & H\&N                       & -     \\
67 & MSD01\_BrainTumor~\cite{77antonelli2022medical}                                                                & MR                 & 3D        & H\&N, $\star$                              & \textcolor{green}{\ding{52}}        \\
68 & MSD02\_Heart~\cite{78simpson2019large}                                                                         & MR                 & 3D        & Pelvic, $\star$                              & \textcolor{green}{\ding{52}}        \\
69 & MSD03\_Liver~\cite{77antonelli2022medical}                                                                     & CT                 & 3D        & Abdomen, $\star$                      & \textcolor{green}{\ding{52}}        \\
70 & MSD05\_Prostate~\cite{77antonelli2022medical}                                                                  & MR                 & 3D        & Pelvic                               & \textcolor{green}{\ding{52}}        \\
71 & MSD06\_Lung~\cite{77antonelli2022medical}                                                                      & CT                 & 3D        & Pelvic, $\star$                              & \textcolor{green}{\ding{52}}        \\
72 & MSD07\_Pancreas~\cite{77antonelli2022medical}                                                                  & CT                 & 3D        & Abdomen, $\star$                      & \textcolor{green}{\ding{52}}        \\
73 & MSD08\_HepaticVessel~\cite{77antonelli2022medical}                                                             & CT                 & 3D        & Abdomen, $\star$                      & \textcolor{green}{\ding{52}}        \\
74 & MSD09\_Spleen~\cite{77antonelli2022medical}                                                                    & CT                 & 3D        & Abdomen                              & \textcolor{green}{\ding{52}}        \\
75 & MSD10\_Colon~\cite{77antonelli2022medical}                                                                     & CT                 & 3D        & Abdomen, $\star$                              & \textcolor{green}{\ding{52}}        \\
76 & MSSEG2016~\cite{85COMMOWICK2021118589}                                                                        & MR                 & 3D        & Other, $\star$                              & -     \\
77 & MSseg08~\cite{msseg2008}                                                                                       & MR                 & 3D        & Thorax, Abdomen, $\star$               & \textcolor{red}{\ding{56}}       \\
78 & MyoPS2020~\cite{87luo2022xmetric,87MyoPS-Net}                                                                  & MR                 & 3D        & Thorax, $\star$                       & -     \\
79 & CT-ORG~\cite{88rister2020ct}                                                                                   & CT                 & 3D        & H\&N, Thorax, Abdomen         & \textcolor{green}{\ding{52}}        \\
80 & PICAI~\cite{89saha2021end}                                                                                     & MR                 & 3D        & Pelvic, $\star$                       & \textcolor{green}{\ding{52}}        \\
81 & PROMISE09~\cite{bharatha2001evaluation}                                                                        & MR                 & 3D        & Pelvic                               & \textcolor{green}{\ding{52}}        \\
82 & PROMISE12~\cite{91litjens2014evaluation}                                                                       & MR                 & 3D        & Pelvic                               & \textcolor{green}{\ding{52}}        \\
83 & Parse2022~\cite{92luo2023efficient}                                                                            & CT                 & 3D        & Thorax                               &  \textcolor{green}{\ding{52}}     \\ 
84 & PH2~\cite{93mendoncca2013ph}                                                                                   & DS         & 2D        & Other, $\star$                              & -     \\ 
85 & Siim-acr-pneumothorax~\cite{94siim-acr-pneumothorax-segmentation}                                           & X-ray              & 2D        & Thorax, $\star$                              & \textcolor{green}{\ding{52}}        \\  
86 & SAML~\cite{95liu2020saml,95liu2020ms}                                             & MR                 & 3D        & Pelvic                               & \textcolor{green}{\ding{52}}        \\  
87 & PCXA~\cite{96jaeger2013automatic,96candemir2013lung}                            & X-ray              & 2D        & Thorax, $\star$                              & \textcolor{green}{\ding{52}}        \\ 
88 & QUBIQ2020~\cite{pal2021holistic}                                                                               & CT                 & 2D        & H\&N, Pelvic                 & \textcolor{green}{\ding{52}}        \\
89 & SLIVER07~\cite{99heimann2009comparison}                                                                        & CT                 & 3D        & Abdomen                              & -     \\
90 & SegTHOR~\cite{100lambert2020segthor}                                                                           & CT                 & 3D        & Thorax, Abdomen                       & -     \\
91 & StructSeg2019~\cite{structseg2019}                                                                             & CT                 & 3D        & H\&N, Thorax, Abdomen, $\star$ & \textcolor{green}{\ding{52}}        \\
92 & TN-SCUI2020~\cite{zhou2020thyroid}                                                                             & US         & 2D        & H\&N,$\star$                              & \textcolor{red}{\ding{56}}       \\
93 & TotalSegmentator~\cite{Wasserthal_2023}                                                              & CT                 & 3D        & H\&N, Thorax, Abdomen, Pelvic  & \textcolor{green}{\ding{52}}        \\
94 & Nerve~\cite{ultrasound_nerve_segmentation}                                           & US         & 2D        & Other                                & \textcolor{green}{\ding{52}}        \\
95 & VESSEL12~\cite{rudyanto2014comparing}                                                                          & CT                 & 3D        & Thorax, $\star$                       & \textcolor{green}{\ding{52}}        \\
96 & VerSe20~\cite{sekuboyina2021verse,loffler2020vertebral}                                                        & CT                 & 3D        & H\&N, Thorax, Abdomen         & \textcolor{green}{\ding{52}}        \\
97 & VerSe19~\cite{sekuboyina2021verse,loffler2020vertebral}                                                        & CT                 & 3D        & H\&N, Thorax, Abdomen         & \textcolor{green}{\ding{52}}        \\
98 & WMH~\cite{kuijf2019standardized}                                                                               & MR                 & 3D        & H\&N, $\star$                & -     \\
99 & WORD~\cite{luo2022word}                                                                                        & CT                 & 3D        & Thorax, Abdomen                       & \textcolor{green}{\ding{52}}        \\
100 & autoPET~\cite{gatidis2022whole}                                                                                & CT, PET             & 3D        & Pelvic, $\star$                       & \textcolor{green}{\ding{52}}        \\
101 & braimMRI~\cite{braimMRI}                                                                                       & MR                 & 2D        & H\&N, $\star$                & \textcolor{green}{\ding{52}}        \\
102 & iSeg2017~\cite{wang2019benchmark}                                                                              & MR                 & 3D        & H\&N                       & -     \\
103 & iSeg2019~\cite{sun2021multi}                                                                                   & MR                 & 3D        & H\&N                       & -     \\
104 & RIM-ONE~\cite{5999143}                                                                                 & FP & 2D        & Other, $\star$                              & -     \\
105 & BUSI~\cite{al2019dataset}                                                                                      & US         & 2D        & Thorax, $\star$                       & \textcolor{green}{\ding{52}}        \\
106 & KvasirCapsule-SEG~\cite{jha2021nanonet}                                                                        & EGD          & 2D        & Other, $\star$                              & \textcolor{green}{\ding{52}}        \\
107 & JSRT~\cite{Shiraishi2000Development}                                                                           & X-ray              & 2D        & Thorax, $\star$                       & -     \\
108 & SZ-CXR~\cite{8477564}                                                                                          & X-ray              & 2D        & Thorax                               & \textcolor{green}{\ding{52}}        \\
109 & EndoVis2017~\cite{allan20192017}                                                                           & EGD          & 2D        & Other, $\star$                                & \textcolor{green}{\ding{52}}        \\
110 & OCCISC~\cite{7386573,7005499}                     & MS         & 2D        & Other                                & -     \\
111 & Kvasir-SEG~\cite{jha2020kvasir}                                                                                & EGD          & 2D        & Other, $\star$                              & \textcolor{green}{\ding{52}}        \\
112 & ISIC18~\cite{tschandl2019ham10000,codella2019skin}                                                             & DS         & 2D        & Other, $\star$                              & \textcolor{green}{\ding{52}}        \\
113 & ISIC17~\cite{codella2018skin}                                                                                  & DS         & 2D        & Other, $\star$                              & \textcolor{green}{\ding{52}}        \\
114 & ISIC16~\cite{gutman2016skin}                                                                                   & DS         & 2D        & Other, $\star$                              & \textcolor{green}{\ding{52}}        \\ \hline
115 & Private01                                                                                                      & CT         & 3D        & Abdomen                             & \textcolor{red}{\ding{56}}        \\ 
116 & Private02                                                                                                      & CT         & 3D        & H\&N                             & \textcolor{red}{\ding{56}}        \\ 
117 & Private03                                                                                                      & CT         & 3D        & H\&N                             & \textcolor{red}{\ding{56}}        \\ 
118 & Private04                                                                                                      & CT         & 3D        & H\&N                             & \textcolor{red}{\ding{56}}        \\ 
119 & Private05                                                                                                      & CT         & 3D        & H\&N                             & \textcolor{red}{\ding{56}}        \\ 
120 & Private06                                                                                                      & CT         & 3D        & H\&N                             & \textcolor{red}{\ding{56}}        \\ 
121 & Private07                                                                                                      & CT         & 3D        & H\&N                             & \textcolor{red}{\ding{56}}        \\ 
122 & Private08                                                                                                      & CT         & 3D        & H\&N                             & \textcolor{red}{\ding{56}}        \\ 
123 & Private09                                                                                                      & CT         & 3D        & Thorax                             & \textcolor{red}{\ding{56}}        \\ 
124 & Private10                                                                                                      & CT         & 3D        & Thorax                             & \textcolor{red}{\ding{56}}        \\ 
125 & Private11                                                                                                      & CT         & 3D        & H\&N, Thorax, Abdomen, Pelvic      & \textcolor{red}{\ding{56}}        \\ 
126 & Private12                                                                                                      & CT         & 3D        & Thorax                             & \textcolor{red}{\ding{56}}        \\ 
127 & Private13                                                                                                      & CT         & 3D        & Thorax                             & \textcolor{red}{\ding{56}}        \\ 
128 & Private14                                                                                                      & CT         & 3D        & Pelvic                             & \textcolor{red}{\ding{56}}        \\ 
129 & Private15                                                                                                      & CT         & 3D        & H\&N, Thorax, Abdomen, Pelvic      & \textcolor{red}{\ding{56}}        \\ 
130 & Private16                                                                                                      & CT         & 3D        & H\&N                             & \textcolor{red}{\ding{56}}        \\ 
131 & Private17                                                                                                      & CT         & 3D        & H\&N                             & \textcolor{red}{\ding{56}}        \\ 
132 & Private18                                                                                                      & CT         & 3D        & Thorax                             & \textcolor{red}{\ding{56}}        \\ 
133 & Private19                                                                                                      & CT         & 3D        & Abdomen                             & \textcolor{red}{\ding{56}}        \\ 
134 & Private20                                                                                                      & CT         & 3D        & Pelvic                             & \textcolor{red}{\ding{56}}        \\ 
135 & Private21                                                                                                      & CT         & 3D        & Pelvic                             & \textcolor{red}{\ding{56}}        \\ 
136 & Private22                                                                                                      & CT         & 3D        & H\&N                             & \textcolor{red}{\ding{56}}        \\ 
137 & Private23                                                                                                      & CT         & 3D        & H\&N                             & \textcolor{red}{\ding{56}}        \\ 
138 & Private24                                                                                                      & CT         & 3D        & H\&N                             & \textcolor{red}{\ding{56}}        \\ 
139 & Private25                                                                                                      & CT         & 3D        & H\&N                             & \textcolor{red}{\ding{56}}        \\ 
140 & Private26                                                                                                      & CT         & 3D        & Pelvic                             & \textcolor{red}{\ding{56}}        \\ 
\hline
\caption{Detailed information about the SA-Med2D-20M dataset. For the column of dataset name, ``Private\textit{xx}'' indicates a private dataset. For the column of anatomical structure, ``H\&N'' indicates ``Head and Neck'', ``$\star$'' indicates the dataset contains tumor-related category. ``Other'' refers to datasets without detailed anatomical information from the original dataset. For the column of release, a ``\textcolor{green}{\ding{52}}'' indicates that we can distribute the processed data to researchers for further analysis, whereas a '\textcolor{red}{\ding{56}}' indicates datasets that cannot be distributed, and ``-'' indicates that we have not obtained explicit permission for redistribution. } 
\label{tab:all_datasets}
\end{longtable}
\end{center}

\bibliographystyle{plain}
\bibliography{citation}

\end{document}